\documentstyle[pre,aps,epsf,multicol]{revtex}
\begin{document}

\title{Phase transition classes in triplet and quadruplet reaction
diffusion models}
\author{G\'eza \'Odor}
\address{Research Institute for Technical Physics and Materials Science, \\
H-1525 Budapest, P.O.Box 49, Hungary}    
\maketitle

\begin{abstract}
Phase transitions of reaction-diffusion systems with site occupation
restriction and with particle creation that requires $n=3,4$ parents, 
whereas explicit diffusion of single particles ($A$) is present are 
investigated in low dimensions by mean-field approximation and simulations. 
The mean-field approximation
of general $nA\to (n+k)A$, $mA\to(m-l)A$ type of lattice models
is solved and novel kind of critical behavior is pointed out.
In $d=2$ dimensions the $3A\to 4A$, $3A\to 2A$ model exhibits a 
continuous mean-field type of phase transition, that implies $d_c<2$ 
upper critical dimension. For this model in $d=1$ extensive simulations 
support a mean-field type of phase transition with logarithmic corrections
unlike the Park et al.'s recent study (Phys. Rev E {\bf 66}, 025101 (2002)).
On the other hand the $4A\to 5A$, $4A\to 3A$ quadruplet model exhibits a
mean-field type of phase transition with logarithmic corrections in $d=2$,
while quadruplet models in 1d show robust, non-trivial transitions suggesting
$d_c=2$. 
Furthermore I show that a parity conserving model $3A\to 5A$, $2A\to\emptyset$
in $d=1$ has a continuous phase transition with novel kind of
exponents. 
These results are in contradiction with the recently suggested implications 
of a phenomenological, multiplicative noise Langevin equation approach 
and with the simulations on suppressed bosonic systems by Kockelkoren 
and Chat\'e (cond-mat/0208497).
\end{abstract}

\begin{multicols}{2}

\section{Introduction}
Phase transitions in genuine nonequilibrium systems have been investigated
often among reaction-diffusion (RD) type of models exhibiting absorbing states
\cite{Dick-Mar,Hin2000,dok}. In many cases mapping to surface growth, spin
systems or stochastic cellular automata can be done.
The classification of universality classes of second order transitions
is still one of the most important uncompleted task. 
One hopes that symmetries and spatial dimensions are the most significant
ingredients as in equilibrium cases, however it turned out that in many cases
there is a shortage of such factors to explain novel universality classes. 
An important example was being investigated during the past two years
that emerges at phase transitions of binary production systems
\cite{HT97,Carlon99,Hayepcpd,Odo00,HayeDP-ARW,coagcikk,binary,multipcpd,NP01,OSC02} 
(PCPD). In these systems particle production competes with pair annihilation and 
single particle diffusion. If the production wins steady states with finite 
particle density appear in (site restricted) models with hard-core repulsion, 
while in unrestricted (bosonic) models the density diverges. By lowering the 
production/annihilation rate a doublet of absorbing states without symmetries 
emerges. One of such states is completely empty, the other possesses a single 
wandering particle. In case of site restricted systems the transition to 
absorbing states is continuous.

Although the nature of this transition has not completely been settled 
numerically and by field theory yet, an other novel class appearing 
in triplet production systems was proposed very recently 
\cite{PHK02,KC0208497} (TCPD). This reaction-diffusion model differs from the
PCPD that for new particle generation at least three particles have to meet. 
It is important to note that these models do not break the DP 
hypothesis \cite{Jan81,Gras82} ---- according to which {\it in one component systems 
exhibiting continuous phase transitions to single absorbing state 
(without extra symmetry and inhomogeneity or disorder) short ranged 
interactions can generate DP class transition only} ---- 
because they exhibit multiple absorbing states that are not frozen, 
lonely particle(s) may diffuse in them.

A phenomenologically introduced Langevin equation that exhibits real, 
multiplicative noise was suggested \cite{PHK02} to describe the critical 
behavior of reaction-diffusion models of types
\begin{equation}
n A \to (n+1)A, \qquad n A \to j A, \label{reactions}
\end{equation}
(with $j<n$ number of interacting particles) in the form
\begin{equation}
\partial_t\rho(x,t) = a \rho(x,t)^n - \rho(x,t)^{n+1} + D\nabla^2\rho(x,t)
+\zeta(x,t), \label{Langevin}
\end{equation}
with noise correlations
\begin{equation}
<\zeta(x,t)\zeta(x',t')> = \Gamma \rho^{\mu} \delta^d(x-x')\delta(t-t') .
\label{noisecorr}
\end{equation}
The classification of universality classes of nonequilibrium
systems by the exponent $\mu$ of a multiplicative noise in the Langevin 
equation was suggested some time ago by Grinstein et al. \cite{GMTMN}.
However it turned out that there may not be corresponding particle systems
to real multiplicative noise cases \cite{HT97} and an imaginary part 
appears as well if one derives the Langevin equation of a RD system 
starting from the Master equation in a proper way. 
This observation led Howard and T\"auber to investigate systems with
complex noise appearing in binary production models.
Unfortunately the cases with and without occupation number restriction
turned out to be different in $d=1$, although in $d=2$ this difference 
was found to disappear at criticality and below \cite{OSC02}. 

By rescaling eq.(\ref{Langevin}) one can get the corresponding 
mean-field critical exponents
\begin{equation}
\beta^{MF}=1, \quad \nu_{\perp}^{MF}=n/2, \quad \nu_{||}^{MF}=n.
\end{equation}
The authors of \cite{PHK02} expect that the noise exponent should be in
the range
\begin{equation}
1\le\mu\le n,
\end{equation}
hence by simple power counting the upper critical dimension should be
\begin{equation}
d_c=2+\frac{4-2\mu}{n} .
\end{equation}
This implies for a triplet processes: $4/3\le d_c \le 8/3$ and for a
quadruplet ($n=4$) processes: $1\le d_c \le 5/2$.

Very recently Kockelkoren and Chat\'e introduced stochastic cellular automata
(SCA) versions of general $nA\to (n+k)A$, $mA\to(m-l)A$ type of models 
\cite{KC0208497},
where multiple particle creation on a given site is suppressed by an 
exponentially decreasing creation probability ($p^{N/2}$) of the particle 
number.
They claim that their simulation results in 1d are in agreement with the
fully occupation number restriction counterparts and set up a general table 
of universality classes, where as the function of $n$ and $m$ only 4 classes 
exist, namely the directed percolation class \cite{Jan81,Gras82}, the parity 
conserving class  \cite{Cardy-Tauber}, the PCPD and TCPD classes.

In any case the heuristic Langevin equation with real noise assumption 
for RD models \cite{PHK02,KC0208497} should be proven for $n>1$. 
Furthermore in low dimensions topological constraints may cause different 
critical behavior with and without occupation number restriction
\cite{OdMe02}.
Note that in case of binary production models it had not been clear at all 
if the $d_c=2$ prediction of the bosonic field theory had also been true for 
site restricted systems until the numerical confirmation of \cite{OSC02}.
In this paper I show simulation results for lattice models with restricted
site occupancy in $d=1,2$ with the aim of locating the upper critical 
dimensions and checking claims the of refs.\cite{PHK02,KC0208497} about 
possible new universality classes.

\section{Mean-field considerations}

In this section I discuss the mean-field equation that can be
set up for site restricted lattice models with general 
microscopic processes of the form
\begin{equation}
n A \stackrel{\sigma}{\to} (n+k)A, 
\qquad m A \stackrel{\lambda}{\to} (m-l) A, \label{genreactions}
\end{equation}
with $n>1$, $m>1$, $k>0$, $l>0$ and $m-l\ge 0$. Note that this formulation 
is different from that of eq.(\ref{Langevin}) that is valid for coarse 
grained, continuous bosonic description of these reaction-diffusion systems.
In this case the diffusion drops out and one can neglect the noise,
hence the the competition of creation 
(with probability $o\sigma$) and annihilation or coagulation 
(parametrized with probability $\lambda=1-\sigma$) is left behind
\begin{equation}
\frac {\partial\rho}{\partial t}= 
a k \sigma \rho^n (1-\rho)^k - a l (1-\sigma) \rho^m, \label{MFeq}
\end{equation}
where $\rho$ denotes the site occupancy probability and
$a$ is a dimension dependent coordination number. 
Each empty site has a probability (1-$\rho$) in mean-field approximation,
hence the need for $k$ empty sites at a creation brings in a $(1-\rho)^k$
probability factor. By expanding $(1-\rho)^k$ and keeping the lowest 
order contribution one can see 
that for site restricted lattice systems a $\rho^{n+1}$-th order term appears 
automatically with negative coefficient that regulates eq.(\ref{MFeq}). 
The steady state solution
can be found analytically in many cases and may result in different,
continuous or discontinuous phase transitions. Here I split the discussion of
the solutions to three parts: (a) $n=m$, (b) $n>m$ and (c) $n<m$.
In the inactive phases one expects a dynamical behavior described by 
the $mA\to\emptyset$ process, for which $\rho\propto t^{1/(m-1)}$ is
known \cite{Cardy-Tauber}.

\subsection{The $n=m$ symmetric case}

The steady state solution in this case can be obtained by solving
\begin{equation}
k \sigma (1-\rho)^k =  l (1-\sigma),
\end{equation}
where the trivial ($\rho=0$) solution has been factored out. For the active
phase one gets
\begin{equation}
\rho = 1 - \left [ \frac{l}{k} \frac{1-\sigma}{\sigma} \right]^{1/k} ,
\end{equation}
which vanishes at $\sigma_c=\frac{l}{k+l}$ with the leading order singularity
\begin{equation}
\rho \propto |\sigma-\sigma_c|^{\beta^{MF}},
\end{equation}
and order parameter exponent exponent $\beta^{MF}=1$. At the critical point
the time dependent behavior is described by
\begin{equation}
\frac {\partial\rho}{\partial t}= -2ak^2\rho^{n+1} + O(\rho^{n+2}) .
\end{equation}
that gives a leading order power-law solution
\begin{equation}
\rho \propto t^{-1/n}
\end{equation}
hence $\alpha_{MF}=\beta^{MF}/\nu^{MF}_{||}=1/n$. 
This was obtained from bosonic, coarse grained formulation in 
\cite{PHK02} too.

\subsection{The $n>m$ case} \label{MFnmsect}

In this case besides the $\rho=0$ absorbing state solution we can get an 
active state if 
\begin{equation}
k\sigma\rho^{n-m}(1-\rho)^k = l(1-\sigma) \label{MFeqnm} 
\end{equation}
is satisfied. Both sides are linear functions of $\sigma$ such that for
$\sigma\to 0$ only the $\rho=0$ is a solution. The left hand side is a
convex function of $\rho$ (from above) with zeros at $\rho=0$ and
$\rho=1$. 
\begin{figure}
\begin{center}
\epsfxsize=80mm
\centerline{\epsffile{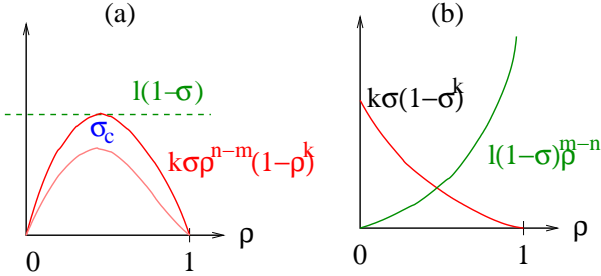}}
\vspace*{4mm}
\caption{Steady state mean-field solution for (a) $n>m$ and (b) $n<m$ cases}
\label{MFabr}
\end{center}
\end{figure}
Therefore by increasing $\sigma$ from zero the left hand side meets 
the right hand side at $\sigma_c$, $\rho_c>0$ (See Fig.\ref{MFabr}(a)).
If this solution is stable a first order transition takes place
in the system.
Note that in higher order cluster mean-field solutions, where the diffusion
can play a role the transition may turn into continuous one 
\cite{boccikk,OdSzo,meorcikk}, therefore it is important to check the type of
transition for $d\ge d_c$. In Section \ref{2dsimunm} I shall confirm
the first orderedness of such transitions for two models in 2d.

\subsection{The $n<m$ case}

By factoring out the trivial $\rho=0$ solution we are faced with
the general condition for a steady state
\begin{equation}
k\sigma(1-\rho)^k = l (1-\sigma)\rho^{m-n} .
\end{equation}
One can easily check that in this case the critical point is at
$\sigma_c=0$ (see Fig.\ref{MFabr}(b))
and here the density decays with $\alpha^{MF}=1/(m-1)$
as in case of the $n=1$ branching and $m=l$ annihilating models
showed by Cardy and T\"auber \cite{Cardy-Tauber} (BkARW classes).
However the steady state solution for $n>1$ gives different $\beta$
exponents than those of BkARW classes, namely $\beta^{MF}=1/(m-n)$.
This imply novel kind of critical behavior in low dimensions, 
that should be a subject of further investigations \cite{brazcikk}.
\begin{figure}
\begin{center}
\epsfxsize=70mm
\centerline{\epsffile{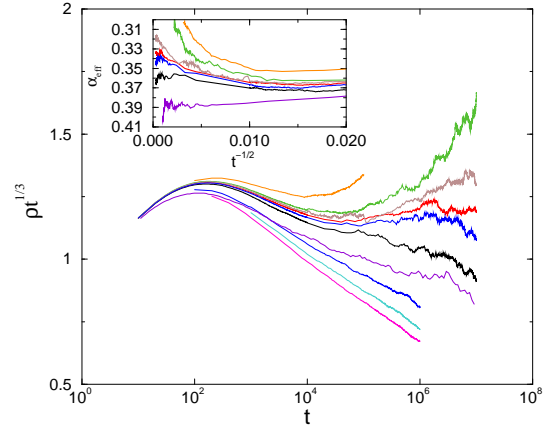}}
\caption{Density times $t^{1/3}$ in the two dimensional 
$3A\to 4A$, $3A\to 2A$ model for $D=0.5$ 
and $p=0.496$, $0.4965$, $0.4967$, $0.4968$, $0.497$, $0.498$, 
$0.4985$ $0.499$ (top to bottom curves). The insert shows the 
corresponding local slopes.}
\label{3A4A-3A2A_5}
\end{center}
\end{figure}

\section{Simulations in two dimensions} \label{2dsimu}

Two dimensional simulations were performed on $L=400-1000$ linear sized 
lattices with periodic boundary conditions. One Monte Carlo step (MCS)
--- corresponding to $dt=1/P$ (where $P$ is the number of particles) ---
is built up from the following processes. A particle and a random
number $x \in (0,1)$ are selected randomly; if $x<D$ a site exchange is 
attempted with one of the randomly selected empty nearest neighbors (nn); 
if $x\ge D$ $k$ number of new particles are created with probability $(1-p)$
at randomly selected empty nn sites provided the number of nn particles was
greater than or equal $n$; or if $x\ge D$ $l$ number of particles are 
removed with probability $p$ (taking into account the $m-l\ge 0$ 
condition as well). 
The simulations were started from fully occupied lattices and
the particle density decay was measured up to $10^6-10^8$ MCS.

\subsection{The $3A\to 4A$, $3A\to 2A$ symmetric triplet model}

First I checked the dynamic behavior in the inactive phase for $D=0.5$
diffusion rate. At $p=0.9$ one can see the appearance of the mean-field
behavior $\rho(t)\propto t^{-1/2}$ following $2\times 10^6$ MCS. 
By decreasing $p$ this scaling sets in later and later times.
As Fig.\ref{3A4A-3A2A_5} shows for $L=1000$ systems with $t_{max}=10^7$
MCS curves with $p\le 0.4965$ veer up 
-- corresponding to the active phase --- while curves with 
$p\ge 0.497$ veer down -- corresponding to the absorbing state.
From the $\rho(t)$ data I determined the effective exponents 
(the local slopes) defined as
\begin{equation}
\alpha_{eff}(t) = {- \ln \left[ \rho(t) / \rho(t/m) \right] 
\over \ln(m)} \label{slopes}
\end{equation}
(where I used $m=4$).
The critical point is estimated at $p=0.4967(2)$ with $\alpha=0.33(1)$
(for local slopes see insert of Fig.\ref{3A4A-3A2A_5}). 
This value agrees with the mean-field value $\alpha^{MF}=1/3$.
\begin{figure}
\epsfxsize=70mm
\epsffile{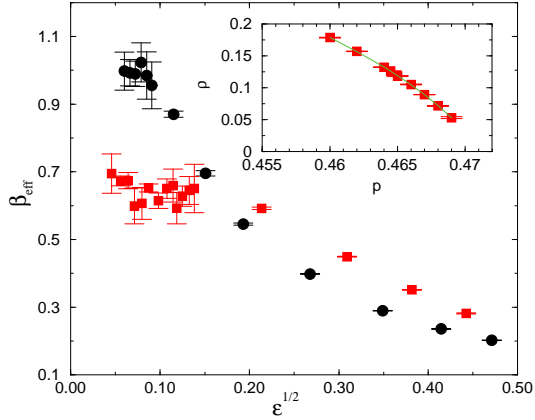}
\vspace{4mm}
\caption{Effective order parameter exponent results in 2d. 
Bullets correspond to the $3A\to 4A$, $3A\to 2A$ model; squares to the 
$4A\to 5A$, $4A\to 3A$ model at $D=0.5$. The insert shows the logarithmic
fitting for the $4A\to 5A$, $4A\to 3A$ model.}
\label{beta_5}
\end{figure}
Density decays for several $p$-s in the active phase 
($0.003 < \epsilon = |p_c-p| < 0.3$) were followed on logarithmic time 
scales and averaging was done over $\sim 100$ independent runs in a time window, 
which exceeds the level-off time by a decade.
The steady state density in the active phase at a critical phase
transition is expected to scale as
\begin{equation}
\rho(\infty,p) \propto |p-p_c|^{\beta} \ .
\end{equation}
Using the local slopes method one can get a precise estimate for
$\beta$ as well as for the corrections to scaling
\begin{equation}
\beta_{eff}(\epsilon_i) = \frac {\ln \rho(\infty,\epsilon_i) -
\ln \rho(\infty,\epsilon_{i-1})} {\ln(\epsilon_i) - \ln(\epsilon_{i-1})} \ \ ,
\label{beff}
\end{equation}
where I used the $p_c$ value determined before. 
One can see on Fig.\ref{beta_5} that the effective exponent for 
$\epsilon>0.005$  exhibits a correction to scaling (inclined line)
and tends to $\lim_{\epsilon\to 0}\beta = 1.0(1)$, which agrees with the 
mean-field value again. By neither the $\alpha$, nor the
$\beta$ exponent can one observe logarithmic corrections suggesting $d_c<2$.

The density decay simulations were repeated at $D=0.2$, where
the critical point was found at $p_c=0.4795(1)$ with mean-field like 
$\alpha$ exponent again.

\begin{figure}
\begin{center}
\epsfxsize=70mm
\centerline{\epsffile{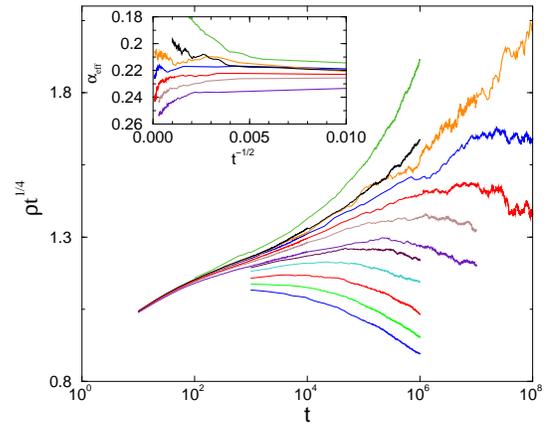}}
\caption{Density times $t^{1/4}$ in the two dimensional 
$4A\to 5A$, $4A\to 3A$ model for $D=0.5$ 
and $p=0.469$, $0.47$, $0.4792$, $0.4705$, $0.471$, $0.4715$, 
$0.4725$, $0.473$, $0.474$, $0.476$, $0.478$, $0.48$ (top to bottom curves).
The insert shows the corresponding local slopes.}
\label{4A5A-4A3A_5}
\end{center}
\end{figure}

\subsection{The $4A\to 5A$, $4A\to 3A$ symmetric quadruplet model}

Here simulations are much slower than in case of the triplet model, 
hence systems with linear size $L=400$ could be investigated. 
First I checked the dynamic behavior in the inactive phase for $D=0.5$.
At $p=0.9$ a mean-field type of decay $\rho(t)\propto t^{-1/3}$ can be
observed following $10^6$ MCS.
As one can see on Fig.\ref{4A5A-4A3A_5} for $p < 0.4702$ the density decay
curves veer up, while for $p\ge 0.4705$ they veer down. The estimated critical
point is $p_c\simeq 0.4703(1)$. The effective exponent at $p_c$ extrapolates
to $\alpha\simeq 0.215(5)$. As one can see on this graph the separatrix
(critical) curve exhibits a linear shape on the $\rho(t)t^{1/4}$ - $\ln(t)$
scale suggesting logarithmic corrections to scaling. Similarly the effective
exponents of $\beta$ seem to extrapolate to $\beta\simeq 0.71(5)$ 
(Fig.\ref{beta_5}) that is very far from the mean-field value $\beta^{MF}=1$.
To check the possibility that a logarithmic correction can result in 
mean-field exponents the fitting with the lowest order correction
\begin{equation}
\rho(\infty,p) = \left[ (p_c-p) (a+b\ln(p_c-p)) \right]^{\beta}
\label{blogfit}
\end{equation}
has been applied for the steady state $\rho(\infty,p)$ data. I used the
non-linear fitting of the ``xmgr'' graphical package with a relative 
error in the sum of squares with at most 0.0001. This resulted in 
$\beta=1.01$ at $p=0.471$ ($a=-10.8$, $b=-6.05$) 
(see insert of Fig. \ref{beta_5}).
This result in agreement with the dynamical scaling conclusion may 
support that the upper critical dimension for quadruplet models is $d_c=2$.
To get more solid results further, very extensive simulations would be
necessary that is beyond the scope of this study.  
In any case clear mean-field behavior can't be concluded.

\begin{figure}
\begin{center}
\epsfxsize=70mm
\centerline{\epsffile{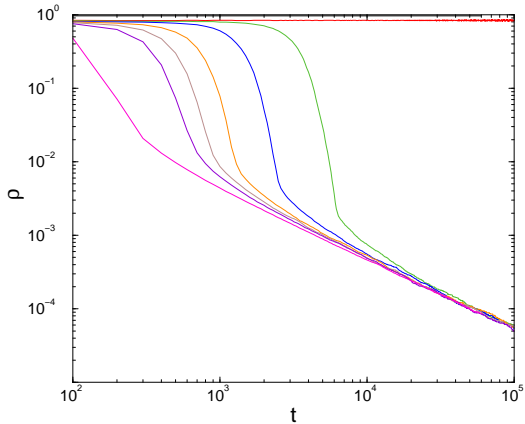}}
\caption{Density decay in the two dimensional $4A\to 5A$, $2A\to\emptyset$
model for $D=0.5$ and $p=0.05$, $0.119$, $0.121$, $0.122$, $0.123$, $0.124$, 
$0.125$, $0.13$ (top to bottom curves) with system sizes $L=400$.}
\label{4A5A-2A0_2d}
\end{center}
\end{figure}

\subsection{$3A\to 5A$, $2A\to \emptyset$ and
            $4A\to 5A$, $2A\to\emptyset$ hybrid models} \label{2dsimunm}

One can find two regions in the density decay behavior by varying $p$
in both models.
For $p<p_c$ steady state values are reached quickly while for $p>p_c$ a
rapid (faster than power-law) initial density decay crosses over to
$\rho \propto t^{-1}$. This is in agreement with the mean-field behavior
of the $2A\to\emptyset$ process in 1d \cite{Lee} dominating in the inactive
phase.
For the $4A\to 5A$, $2A\to\emptyset$ quadruplet production model 
this threshold is at $p_c=0.119(1)$ (see Fig.\ref{4A5A-2A0_2d}) where
an abrupt jump is observable from $\rho(\infty)=0.833$ to $\rho(\infty)=0$.
In case of the $3A\to 5A$, $2A\to \emptyset$ triplet production model the
threshold is at $p_c=0.220(1)$ with a jump from $\rho(\infty)=0.45$ to zero.

In neither cases do we see dynamical scaling at the transition.
These results are in agreement with the first order transition of the
mean-field approximations given in Sect.\ref{MFnmsect} for $n>m$.

\section{Simulations in one dimension} \label{1dsimu}
The simulations in one dimension were carried out on $L=10^5$ sized
systems with periodic boundary conditions. The initial states were again
fully occupied lattices, and the density of particles is followed up to
$10^6$ MCS. An elementary MCS consists of the following processes:
\begin{description}
\item[(a)] $A\emptyset\leftrightarrow\emptyset A$ with probability D,
\item[(b)] $m A \to (m-l)A$ with probability $p(1-D)$,
\item[(c)] $n A \to (n+k)A$ with probability $(1-p)(1-D)$,
\end{description}
such that the reactions were allowed on the left or right side of the 
selected particle strings.

\subsection{$3A\to 4A$, $3A\to 2A$ and $3A\to 6A$, $3A\to\emptyset$
symmetric triplet models}

\begin{figure}
\epsfxsize=70mm
\epsffile{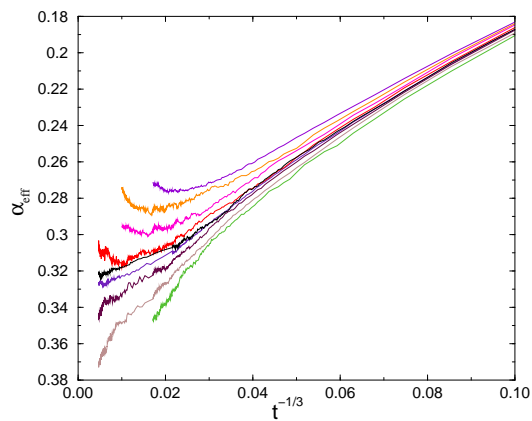}
\vspace{2mm}
\caption{Local slopes of the density decay in the 1d
$3A\to 4A$, $3A\to 4A$ model at $D=0.1$.
Different curves correspond to $p=0.3$, $0.301$, $0.3015$, $0.302$ 
$0.3025$, $0.3027$, $0.303$, $0.3035$, $0.304$ and $0.3045$ 
(from top to bottom).}
\label{3A4A-3A2A_1}
\end{figure}

The $3A\to 4A$, $3A\to 2A$ site restricted model in 1d was simulated by 
Park et al.
\cite{PHK02} for small systems up to $10^6$ MCS. They concluded to find a
novel kind of phase transition with the order parameter exponents 
$\alpha=0.32(1)$ and $\beta=0.78(3)$. For the restricted bosonic version
of this model large scale simulations gave $\alpha=0.27(1)$ and 
$\beta=0.90(5)$ \cite{KC0208497}. Note that since reactive and diffusive
sectors arise in this model like in PCPD, diffusion dependence or
corrections to scaling may hamper to see real asymptotic behavior
\cite{Odo00,DM0207720,Odo02}.
Here I show extended simulation results for the strictly site restricted
lattice model model with $t_{max}=10^7$ MCS at diffusion rate 
$D=0.1$. At the critical point the $\alpha_{eff}(t)$ curve
exhibits a straight line shape for $t\to\infty$, while in 
sub(super)-critical cases $\alpha_{eff}(t)$ curves veer down(up) 
respectively. 
As one can see on Fig.\ref{3A4A-3A2A_1} following a long relaxation 
$p\le 0.3032$ curves veer up, while $p\ge 0.3035$ curves veer down in the
$t\to\infty$ limit. 
From this one can estimate $p=p_c\simeq 0.3033(1)$ with $\alpha=0.33(1)$ 
in agreement with the results of \cite{PHK02}. 

\begin{figure}
\epsfxsize=70mm
\epsffile{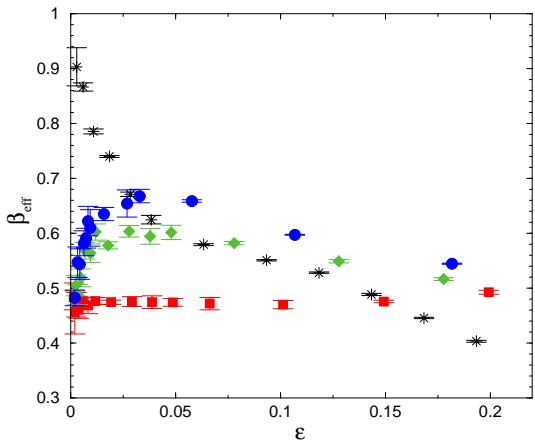}
\vspace{4mm}
\caption{Effective order parameter exponent results in 1d. 
Stars correspond to $3A\to 4A$, $3A\to 2A$ model; bullets to
$4A\to 5A$, $2A\to\emptyset$ model at $D=0.2$; squares to 
$4A\to 5A$, $2A\to\emptyset$ model at $D=0.8$; diamonds to
$4A\to 5A$, $4A\to\emptyset$ model at $D=0.3$.}
\label{beta3A4A-3A2A_1}
\end{figure}

By analyzing super-critical, steady state densities with the local slopes 
method one can read-off: 
$\beta_{eff}\to\beta \simeq 0.95(5)$ (see Fig.\ref{beta3A4A-3A2A_1}), 
which is higher than the results of \cite{PHK02} and \cite{KC0208497}.

However one should be careful and check diffusion dependence and
corrections to scaling especially because these critical exponent 
estimates are quite close to the mean-field values 
($\alpha^{MF}=1/3$, $\beta^{MF}=1$) and as it was shown in 
Sect. \ref{2dsimu} $d_c<2$. Since the $p=0.303$ and $p=0.3035$ 
$\rho(t)$ curves show clear curvature for large times 
the $0.303 < p_c < 0.3035$ conditions seems to be inevitable.
I tried to fit the steady state data in the $0.303\le p\le 0.3035$ region 
by the logarithmic correction form (\ref{blogfit}) and obtained
$\beta=1.07(10)$ at $p_c=0.3032$ that agrees with the mean-field value
and implies $d_c=1$. 

Just considering mean-field results, according to which $k$ does not play a 
role for $n=m$ models one may expect that the $3A\to 6A$, 
$3A\to\emptyset$ triplet creation model exhibits the same kind of 
transition as the $3A\to 4A$, $3A\to 2A$ model. Indeed Kockelkoren and 
Chat\'e's simulations show this \cite{KC0208497}.
However doing lattice simulations with site restrictions it turned out that the 
$3A\to 6A$ creation was so effective that it shifted the transition to the 
zero production limit ($p=1$) where the $3A\to\emptyset$ process in 1d is 
known to decay as $\rho \propto (\ln(t)/t)^{\frac12}$ \cite{Cardy-Tauber}. 
Off-critical simulations showed that $\beta=0.33(1)$, meaning that this
transition belongs to the BkARW mean-field class. 
On the other hand there may be other realizations of this model,
where the transition reported by Kockelkoren and Chat\'e is accessible.

\subsection{The $3A\to 5A$, $2A\to \emptyset$ model}

It has been established that in $n=1$, $m=l=2$ and even $k$ -- so called
even number of offspringed branching and annihilating models (BARWe) 
-- the parity conserving (PC) class continuous phase transition emerges
\cite{Cardy-Tauber,Taka,IJen94}. This class has also been observed in models
exhibiting $Z_2$ symmetric absorbing states, where the domain walls separating
ordered phases follow BARWe dynamics \cite{Gras84,nekim,Park94,Bas,Hin97}.
This class was originally called parity conserving, owing to the 
conservation law that made it different from the robust DP class. 
However it turned out that in $Z_2$ symmetric models this conservation 
is not enough \cite{Parkh,meod96,Hin97,odme98}. Furthermore in binary 
spreading models this conservation was found to be irrelevant 
\cite{binary,OSC02}. Therefore it is still an open question whether 
parity conservation is relevant in other models than in BARW types. 
\begin{figure}
\begin{center}
\epsfxsize=70mm
\centerline{\epsffile{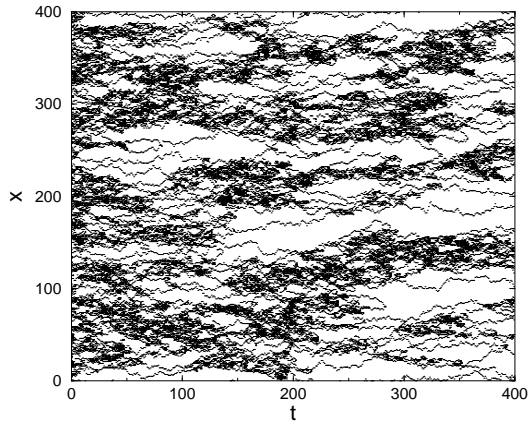}}
\caption{Spatio-temporal evolution of the critical, 1+1 d diffusive 
$3A\to 5A$, $2A\to \emptyset$ model ($D=0.8$).}
\label{3A5A_2A0_8rajz}
\end{center}
\end{figure}
I investigated the phase transition of the triplet production 
$3A\to 5A$, $2A\to \emptyset$ model (with explicit particle diffusion)
possessing parity conservation.
As I showed in Sect. \ref{2dsimunm} in two dimensions this system exhibits 
a first order transition in agreement with the mean-field results. 
This first order mean-field behavior does not give a direct hint on the 
type of phase transition in 1d. Kockelkoren and Chat\'e's simulations on
the one dimensional, suppressed bosonic cellular automaton version of this 
model shows simple DP class density decay \cite{KC0208497}. 
However if we consider the space-time evolution we see very non-DP like 
spatio-temporal pattern (see Fig.\ref{3A5A_2A0_8rajz}).
This pattern resembles much more to those of the PCPD class models, where
compact domains separated by clouds of lonely wandering particles occur.
Of course such qualitative judgment on the universal behavior is not enough
but has been found to be quite successful in case of binary production
systems \cite{HayeDP-ARW,multipcpd}. 

\begin{figure}[h]
\epsfxsize=70mm
\epsffile{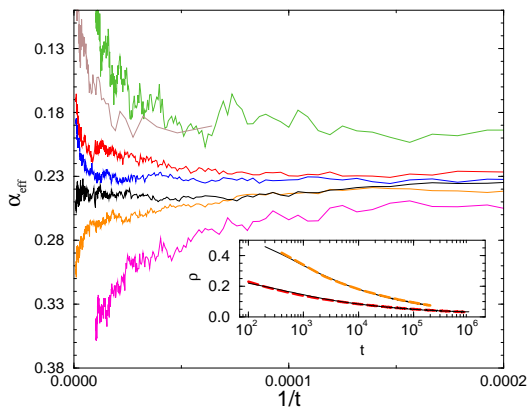}
\vspace{2mm}
\caption{Local slopes of the density decay in the 1d
$3A\to 5A$, $2A\to \emptyset$ model for $D=0.8$.
Different curves correspond to $p=0.46$, $0.461$, $0.462$, $0.4625$, $0.4627$, 
$0.463$, $0.46325$, $0.464$ and $0.465$ (from top to bottom).}
\label{3A5A-2A0_8}
\end{figure}

The density decay simulations at
$D=0.8$ and $D=0.2$ have been analyzed by the local slopes method see 
Fig.\ref{3A5A-2A0_8}. 
At $D=0.8$ the critical point is estimated at $p_c^H=0.4629(3)$
and the corresponding effective exponent tends to $\alpha^H = 0.24(1)$.
At $D=0.2$-s the critical point is at $p_c^L = 0.2240(3)$, and the local 
exponent seems to extrapolate to $\alpha^L = 0.28(1)$. 
Such small difference between the high and low $D$ results can
also be observed by analyzing the steady state results:
$\beta^H=0.43(3)$ versus $\beta^L=0.63(3)$.  
These exponent estimates are far from the 1+1 d DP values 
($\alpha=0.159464(6)$, $\beta=0.276486(8)$ \cite{IJen99}),
hence the claim of Kockelkoren and Chat\'e for the critical behavior
of $n>m$ models is questionable.
On the other hand the diffusion dependence of the critical exponents
is a challenge and has been observed in the binary production 
PCPD model \cite{Odo00}. In \cite{Odo02} it was shown that assuming
logarithmic corrections to scaling -- that is quite common in 1d models --
a single universality class can be supported numerically. Therefore here
again I have investigated the possibility of the collapse of the high and 
low $D$ exponents. Assuming the same kind of logarithmic correction forms 
as in \cite{Odo02}
\begin{equation}
\left[ [ a + b\ln(t) ] / t \right] ^{\alpha} \label{betalogform}
\end{equation}
I have found a consistent set of exponents both for $D=0.2$ and $D=0.8$
(see Table \ref{3A5Alogtable} and insert of Fig.\ref{3A5A-2A0_8}).
For the data analysis I used non-linear fitting of the program ``xmgr'' 
package, with a relative error in the sum of squares with at most 0.001.
\begin{equation}
\alpha=0.22(1), \qquad \beta=0.60(1) \ ,
\end{equation}
with the critical thresholds: $p_c^H=0.4627(1)$, $p_c^L=0.2240(1)$.
These exponents suggest a distinct universality class from the known
ones \cite{dok}.

\subsection{$4A\to 5A$, $4A\to\emptyset$ and $4A\to 5A$, $2A\to\emptyset$
quadruplet models}

\begin{figure}
\epsfxsize=70mm
\epsffile{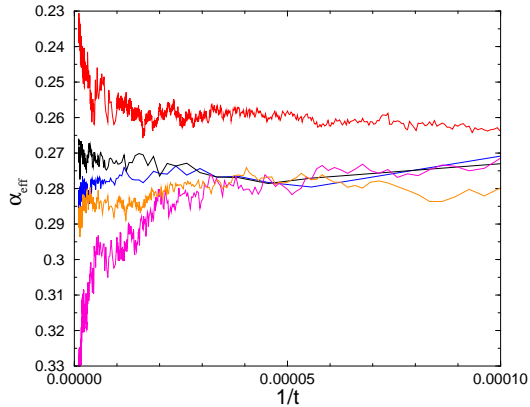}
\vspace{2mm}
\caption{Local slopes of the density decay in the 1d
$4A\to 5A$, $4A\to \emptyset$ model for $D=0.3$.
Different curves correspond to $p=0.904$, $0.9037$, $0.9033$, $0.903$ 
$0.9027$, $0.902$ (from bottom to top).}
\label{4A_3}
\end{figure}

Two dimensional simulations (Sect.\ref{2dsimu}), showed that
for $n=m=4$ symmetric quadruplet models $d_c=2$.
Simulations in the corresponding suppressed bosonic SCA \cite{KC0208497} 
with $n=4$ and $1\le m \le 4$ located the phase transition at zero 
production rate. Here I show that in the one dimensional 
$4A\to 5A$, $4A\to\emptyset$ and $4A\to 5A$, $2A\to\emptyset$ 
site restricted models continuous phase transitions with $p<1$ and with
non-trivial exponents can be found. 
The density decay was followed up to $t=10^6$ MCS and the critical 
point was located by the local-slopes method (see Fig.\ref{4A_3})
at $p=0.9028(1)$ for $D=0.3$.
The corresponding exponent can be estimated as $\alpha=0.27(1)$.
For $D=0.05$ one gets $p_c=0.9605(3)$ with $\alpha=0.28(1)$,
so one can not see diffusion dependence here.
Analyzing off-critical data with the local slopes method
(\ref{beff}) one gets $\beta=0.48(2)$ (see Fig.\ref{beta3A4A-3A2A_1}).
In accordance with these results simulations for the $4A\to 5A$, 
$2A\to\emptyset$ model at $D=0.2$ and $D=0.8$ diffusion rates resulted 
in $p_c(0.2) = 0.53185(5)$, $p_c(0.8) = 0.5742(1)$ with
$\alpha=0.27(1)$ and $\beta=0.48(2)$ exponents
(see Fig.\ref{beta3A4A-3A2A_1}).
As we can see critical exponent data for quadruplet models are robust and 
no diffusion dependence has been found. Furthermore critical space-time plots 
are very similar to that of the PCPD model.

\section{Conclusions}

In this paper I investigated the phase transitions of general $nA\to (n+k)A$, 
$mA\to(m-l)A$ reaction type of models with explicit single particle diffusion 
on occupation number restricted lattices in one and two dimensions.
I showed that mean-field solution for $n=m$ symmetric cases results in
universality classes characterized by the exponents $\alpha=1/n$, $\beta=1$.
I determined the upper critical dimensions for the triplet and quadruplet
cases by simulations. For $n=3$ high precision simulations show mean-field 
type of criticality with logarithmic corrections meaning $d_c=1$. 
This result is in contradiction with the simulations of \cite{PHK02} and 
\cite{KC0208497} and with the analytical form for $d_c(n)$ derived from a 
phenomenological Langevin equation.
In case of my site restricted realization of the one dimensional 
$3A\to 6A$, $3A\to\emptyset$ model the phase transition point is shifted 
to zero production rate and is continuous, BkARW mean-field type. This is
in contradiction with the findings of \cite{KC0208497} for an other
stochastic cellular automaton realization of this model.
For $n=4$ the upper critical dimension was located at $d_c=2$ opening up
the possibility for non-trivial critical behavior in $d=1$.
Indeed two versions of such quadruplet models were shown to exhibit
robust, novel type of critical transition in one dimension.

For $n>m$ the mean-field approximation gives first order transition
that was observed by simulations for two ($n=3,4$) models in $d=2$. 
On the other hand numerical evidence was given that the parity conserving 
model $3A\to 5A$, $2A\to \emptyset$ in 1d exhibits a non-PC type of 
critical behavior with logarithmic corrections by varying the diffusion
rate. This transition does not fit in the universality class scheme 
suggested by \cite{KC0208497}.

Finally I showed that for $n<m$ models the mean-field approximations 
result in new classes featured by $\alpha^{MF}=1/(m-1)$ and 
$\beta^{MF}=1/(m-n)$. Such kind of models should be subject of further 
studies. 

The presented mean-field and simulation results show that the universal
behavior of such low-dimensional reaction-diffusion models is rich and
the table of universality classes given by \cite{KC0208497} is not valid
for 1d, fully site restricted systems. Perhaps the strict site restriction 
plays an important role that causes the differences. 
Field theoretical (possibly fermionic) treatment that starts from the 
master equation should be set up to determine at least the analytical 
form of $d_c(n)$ for $n=m>2$ models.

\vskip 0.5cm

\noindent
{\bf Acknowledgements:}\\
I thank H. Chat\'e, H. Hinrichsen and U. T\"auber for useful communications 
and N. Menyh\'ard for her comments to the manuscript.
Support from Hungarian research funds OTKA (Grant No. T-25286), Bolyai
(Grant No. BO/00142/99) and IKTA (Project No. 00111/2000) is acknowledged.
The simulations were performed on the parallel cluster of SZTAKI and on the
supercomputer of NIIF Hungary within the scope of the DEMOGRID project.

\begin{table}[h]
\caption{Logarithmic fitting results by the form (\ref{betalogform})
for the one dimensional $3A\to 5A$, $2A\to \emptyset$ model.
\label{3A5Alogtable}}
\begin{tabular}{|c|c|c|c|c|} \\
\hline
$D$   & $p_c$     & $a$     & $b$                & $\alpha$ \\
\hline
$0.2$ & $0.4627$  & $0.115$ & $3.5\times10^{-5}$ & $0.217$ \\
$0.8$ & $0.22405$ & $14.12$ & $-1.001$           & $0.224$ \\
\end{tabular}
\end{table}

\end{multicols}

\begin{references}

\bibitem{Dick-Mar} For references see : J.~Marro and R.~Dickman,
\newblock {\em Nonequilibrium phase transitions in lattice models},
\newblock Cambridge University Press, Cambridge, 1999.
\bibitem{Hin2000} For a review see : H.~Hinrichsen, 
Adv. Phys. {\bf 49}, 815 (2000).
\bibitem{dok} For a recent review see : G. \'Odor,
{\em Phase transition universality classes of classical, 
nonequilibrium systems}, cond-mat/0205644.

\bibitem{HT97} M.~J. Howard and U.~C. T{\"a}uber,  
{J. Phys.} {\bf A 30}, 7721 (1997).
\bibitem{Carlon99} E.~Carlon, M.~Henkel and U.~Schollw{\"o}ck,
Phys. Rev. E {\bf 63}, 036101-1 (2001).
\bibitem{Hayepcpd} H.~Hinrichsen, Phys. Rev. E {\bf 63}, 036102-1 (2001).
\bibitem{Odo00} G. \'Odor , Phys. Rev. {\bf E62}, R3027 (2000).
\bibitem{HayeDP-ARW} H.~Hinrichsen, Physica A {\bf 291}, 275-286 (2001).
\bibitem{coagcikk} G. \'Odor , Phys. Rev. E {\bf 63}, 067104 (2001).
\bibitem{binary} K. Park, H. Hinrichsen, and In-mook Kim, Phys. Rev. E
{\bf 63}, 065103(R) (2001).
\bibitem{multipcpd} G. \'Odor, Phys. Rev. E {\bf 65}, 026121 (2002).
\bibitem{NP01} J. D. Noh and H. Park, cond-mat/0109516.
\bibitem{OSC02} G. \'Odor, M. A. Santos, M. C. Marques,
Phys. Rev. E {\bf 65}, 056113 (2002).

\bibitem{PHK02} K. Park, H. Hinrichsen and I. Kim, Phys. Rev. E {\bf 66},
025101 (2002).
\bibitem{KC0208497} J. Kockelkoren and H. Chat\'e, cond-mat/0208497.

\bibitem{Jan81} H. K. Janssen, Z. Phys. B {\bf 42}, 151 (1981).
\bibitem{Gras82} P. Grassberger, Z. Phys. B {\bf 47}, 365 (1982).

\bibitem{GMTMN} Grinstein G., Mu\~noz M.A., and Tu Y.,
                Phys. Rev. Lett. {\bf 76}, 4376 (1996);
                Tu Y., Grinstein G., and Mu\~noz M.A.
                Phys. Rev. Lett. {\bf 78}, 274 (1997).

\bibitem{Cardy-Tauber} J. L. Cardy and U. C. T\"auber,
Phys. Rev. Lett. {\bf 77}, 4780 (1996); J. L. Cardy and U. C. T\"auber,
J. Stat. Phys. {\bf 90}, 1 (1998).

\bibitem{OdMe02} G. \'Odor and N. Menyh\'ard, Physica D {\bf 168}, 305 (2002).
\bibitem{brazcikk} G. \'Odor, cond-mat/0304023.

\bibitem{boccikk} G. \'Odor, N. Boccara, G. Szab\'o,
Phys. Rev.  E {\bf 48}, 3168 (1993).
\bibitem{OdSzo} G. \'Odor and A. Szolnoki,
Phys. Rev. E {\bf 53}, 2231 (1996).
\bibitem{meorcikk} N. Menyh\'ard and G. \'Odor,
J. Phys. A. {\bf 28}, 4505 (1995).
\bibitem{Lee} B. P. Lee, J. Phys. A {\bf 27}, 2633 (1994).


\bibitem{woh} F. van Wijland, K. Oerding and H. J. Hilhorst, Physica A
{\bf 251}, 179 (1998).


\bibitem{Taka} H. Takayasu and A. Yu. Tretyakov, Phys.\ Rev.\ Lett.
{\bf 68}, 3060, (1992).
\bibitem{IJen94} I. Jensen, Phys. Rev. E. {\bf 50}, 3623 (1994).
\bibitem{Gras84} P.~Grassberger, F.~Krause and T.~ von der
Twer, J. Phys. A:Math.Gen., {\bf 17}, L105 (1984).
\bibitem{nekim} N.~ Menyh\'ard, J. Phys. A {\bf 27}, 6139 (1994),
\bibitem{Park94} M. H. Kim and H. Park,  Phys.\ Rev.\ Lett.
{\bf 73}, 2579, (1994).
\bibitem{Bas} K. E. Bassler and D. A. Browne,
   Phys. Rev. Lett. {\bf 77}, 4094 (1996).
\bibitem{Parkh} H.Park and H.Park, Physica A  {\bf 221}, 97 (1995).
\bibitem{meod96} N.~ Menyh\'ard and G. ~\'Odor, 
J. Phys. A {\bf 29}, 7739 (1996).
\bibitem{Hin97} H. Hinrichsen, Phys. Rev. {\bf E 55}, 219 (1997).
\bibitem{odme98}G. \'Odor and N. Menyh\'ard, Phys. Rev. E {\bf 57},5168 (1998).

\bibitem{MHUS} M.~Henkel and U.~Schollw{\"o}ck,
J. Phys. A {\bf 34}, 3333 (2001).

\bibitem{PK66} K. Park and I. Kim, Phys. Rev E {\bf 66}, 027106 (2002).
\bibitem{DM0207720} R. Dickman and M. A. F. de Menenzes,
                    Phys. Rev. E {\bf 66}, 045101 (2002).

\bibitem{Odo02} G. \'Odor, cond-mat/0209287.

\bibitem{G82} P. Grassberger, Z. Phys. B {\bf 47}, 365 (1982).

\bibitem{IJen99} I. Jensen, J. Phys. A {\bf 32}, 5233 (1999).

\end{references}
\end{document}